\documentclass[prd,twocolumn,eqsecnum,nofootinbib]{revtex4}
\usepackage{bm} 
\usepackage{amssymb}
\begin{document}
\title{A light-cone gauge for black-hole perturbation theory}
\author{Brent Preston and Eric Poisson}
\affiliation{Department of Physics, University of Guelph, Guelph, 
Ontario, Canada N1G 2W1}
\date{June 20, 2006} 
\begin{abstract} 
The geometrical meaning of the Eddington-Finkelstein coordinates of   
Schwarzschild spacetime is well understood: (i) the advanced-time
coordinate $v$ is constant on incoming light cones that converge
toward $r=0$, (ii) the angles $\theta$ and $\phi$ are constant on the
null generators of each light cone, (iii) the radial coordinate $r$ is
an affine-parameter distance along each generator, and (iv) $r$ is an 
areal radius, in the sense that $4\pi r^2$ is the area of each
two-surface $(v,r) = \mbox{constant}$. The light-cone gauge of
black-hole perturbation theory, which is formulated in this paper, 
places conditions on a perturbation of the Schwarzschild metric that
ensure that properties (i)--(iii) of the coordinates are preserved in
the perturbed spacetime. Property (iv) is lost in general, but it is
retained in exceptional situations that are identified in this
paper. Unlike other popular choices of gauge, the light-cone gauge
produces a perturbed metric that is expressed in a meaningful
coordinate system; this is a considerable asset that greatly
facilitates the task of extracting physical consequences. We
illustrate the use of the light-cone gauge by calculating the metric 
of a black hole immersed in a uniform magnetic field. We construct a
three-parameter family of solutions to the perturbative
Einstein-Maxwell equations and argue that it is applicable to a
broader range of physical situations than the exact, two-parameter 
Schwarzschild-Melvin family.      
\end{abstract} 
\maketitle

\section{Introduction} 

The theory of perturbations of the Schwarzschild spacetime is a
well-developed one \cite{chandrasekhar:83}, and it may seem surprising
that authors are still writing on this venerable topic almost fifty 
years after its inception in the work of Regge and Wheeler
\cite{regge-wheeler:57} (see also Refs.~\cite{vishveshwara:70,
zerilli:70}). This is (at least partially) explained by the fact that
the field has witnessed a resurgence of sorts in the last several
years, motivated by new applications that include the gravitational 
self-force problem \cite{poisson:04b, detweiler:05, hikida-etal:05,
drasco-etal:05}, the ``close-limit'' collision of two black holes
\cite{gleiser-etal:00}, and the study of the dynamics of black holes
placed in tidal environments \cite{alvi:00, alvi:01, alvi:03,
yunes-etal:06, yunes-gonzalez:06}. The theory has been presented
in various sophisticated packages \cite{moncrief:74,
gerlach-sengupta:79, sarbach-tiglio:01, gundlach-martingarcia:00, 
clarkson-barrett:03, nagar-rezzolla:05, martel-poisson:05}, and it has
reached what is likely to be its definitive form. 

We wish to make an additional contribution to this body of literature
by formulating a useful, attractive, and simple gauge condition for
black-hole perturbation theory. We believe that this gauge, which we 
call {\it the light-cone gauge}, is preferable in many ways to most
popular gauges, including the oft-used Regge-Wheeler gauge. We believe
that the use of the light-cone gauge will be a great benefit to any
researcher faced with the task of computing and interpreting a
perturbation of the Schwarzschild spacetime.     

The idea is simple. The difficulties of the Schwarzschild coordinates
$(t,r,\theta,\phi)$ across the black-hole horizon are well documented,
and it is well known that the transformation $v = t + r  
+ 2M\ln(r/2M - 1)$ brings the metric to a form that is well-behaved on  
the event horizon. The Eddington-Finkelstein coordinates
$(v,t,\theta,\phi)$ have a clear geometrical meaning. The null
coordinate $v$ (called advanced time) is constant on incoming light
cones that converge toward $r=0$, the angles $\theta$ and $\phi$ are 
constant on the null generators of each light cone $v =
\mbox{constant}$, and $r$ is an affine-parameter distance along each
generator. In addition, $r$ doubles as an areal radius, in the sense
that $4\pi r^2$ is the area of each two-sphere $(v,r) =
\mbox{constant}$.  

{\it The light-cone gauge places conditions on the metric perturbation
that ensure that the geometrical meaning of the coordinates is
preserved in the perturbed spacetime.} The advanced-time coordinate
$v$ therefore continues to label incoming light cones that converge
toward $r=0$, the angles $\theta$ and $\phi$ continue to label the
generators of each light cone, and $r$ continues to be an
affine-parameter distance along each generator. One geometrical aspect
of the coordinates that must generally be given up is the role of $r$
as an areal radius; we shall show, however, that this property also
can be preserved in special circumstances. The light-cone gauge
therefore produces a perturbed metric that is expressed in a
meaningful coordinate system. This is a considerable asset that
greatly facilitates the task of extracting the physical properties of
the spacetime.   

The light-cone gauge is developed in Sec.~II of this paper. The gauge
conditions are introduced in Sec.~II B, after we present in Sec.~II A
a brief review of the Schwarzschild metric in Eddington-Finkelstein
coordinates. In Sec.~II C and D we decompose the metric perturbation
in spherical harmonics and explore the space of gauge transformations
that keep the perturbation within the light-cone gauge. This remaining 
gauge freedom is convenient, as it can be exploited to simplify the
form of the perturbed metric to the fullest extent possible. In
Sec.~II E we determine the conditions under which $r$ retains its 
interpretation as an areal radius. The answer turns out to be simple:
This is possible whenever the component $T^{vv} = T_{rr}$ of the
perturbing energy-momentum tensor vanishes. In Sec.~II F we summarize
our construction and discuss its merits; in particular we compare our
light-cone gauge to the very widely used, but far less compelling,
Regge-Wheeler gauge.  

In Sec.~III we present an illustrating application of the light-cone
gauge for black-hole perturbation theory: We examine a black hole
immersed in a uniform magnetic field, and calculate its metric
accurately through second order in the strength of the magnetic
field. The physical situation is described in Sec.~III A. The magnetic
field and its energy-momentum tensor are computed in Sec.~III B. In
Sec.~III C and D we integrate the equations of black-hole perturbation
theory for this situation. The solution is presented in Sec.~III E,
and in Sec.~III F we examine the structure of the perturbed
horizon. Finally, in Sec.~III G we compare our perturbative solution
to the exact Schwarzschild-Melvin solution \cite{ernst:76,
ernst-wild:76, hiscock:81}, which also describes a magnetized black 
hole. We conclude that the perturbative solution is applicable to a
broader range of physical situations.   

In the Appendix we provide a complete listing of the linearized field
equations in the light-cone gauge. 

Our developments in this paper rely heavily on the recent work of
Martel \& Poisson \cite{martel-poisson:05}, which presents a
covariant and gauge-invariant formalism for black-hole perturbation
theory. They also incorporate key ideas from a companion paper
\cite{preston-poisson:06a} devoted to the construction of light-cone
coordinates centered on a geodesic world line of an arbitrary curved
spacetime. 

We point out that the light-cone gauge constructed here is adapted
specifically to incoming light cones $v = \mbox{constant}$ that
converge toward $r=0$. It would be exceedingly straightforward to
adapt the construction to outgoing light cones $u = \mbox{constant}$
that expand toward $r=\infty$. [In Schwarzschild spacetime, the
retarded time coordinate $u$ is defined by $u = t - r 
- 2M\ln(r/2M - 1)$.] While the incoming light-cone gauge is well
suited to study the properties of the perturbed event horizon, the
outgoing light-cone gauge would be well suited to study the
gravitational radiation escaping toward future null infinity. We
suggest this adaptation as an exercise for the reader. 

Throughout the paper we work in geometrized units $(c = G = 1)$ and
adhere to the conventions of Misner, Thorne, and Wheeler
\cite{MTW:73}.       

\section{Light-cone gauge: definition and properties} 

\subsection{Schwarzschild metric in light-cone coordinates} 

The transformation $v = t + r + 2M\ln(r/2M-1)$ brings the
Schwarzschild metric from its usual form to the Eddington-Finkelstein
form  
\begin{equation} 
ds^2 = -f\, dv^2 + 2\, dvdr + r^2\, d\Omega^2, 
\label{2.1}
\end{equation} 
where
\begin{equation} 
f := 1 - \frac{2M}{r} 
\label{2.2} 
\end{equation} 
and 
\begin{equation} 
d\Omega^2 := \Omega_{AB}\, d\theta^A d\theta^B 
:= d\theta^2 + \sin^2\theta\, d\phi.
\label{2.3}
\end{equation} 
The parameter $M$ is the black-hole mass and $\theta^A =
(\theta^2,\theta^3) = (\theta,\phi)$ are angles that span the
two-spheres $(v,r) = \mbox{constant}$. 

The coordinates $(v,r,\theta,\phi)$ are well-behaved across the event 
horizon, and they possess a clear geometrical meaning. We note first
that the vector   
\begin{equation} 
l_\alpha := -\nabla_\alpha v = (-1,0,0,0) 
\label{2.4}
\end{equation} 
is null, which implies that each surface $v = \mbox{constant}$ is a  
null hypersurface of the Schwarzschild spacetime; these are in fact
incoming light cones that converge toward the black-hole
singularity. The fact that $l_\alpha$ is a gradient implies that
$l^\alpha$ is everywhere tangent to a congruence of null geodesics;
these are affinely parameterized, and they are the generators of each
light cone $v = \mbox{constant}$. Using the metric of Eq.~(\ref{2.1})
to raise indices, we find that    
\begin{equation} 
l^\alpha = (0,-1,0,0).
\label{2.5}
\end{equation} 
This relation implies that $\theta^A = \mbox{constant}$ on the
generators, so that the angles $\theta^A$ can be used as generator
labels. Furthermore, Eq.~(\ref{2.5}) reveals that the affine
parameter on each generator is $-r$. The geometrical meaning of the
coordinates is therefore the following: The null coordinate $v$
(called advanced time) is constant on incoming light cones that
converge toward $r=0$, $\theta^A$ labels the generators of each light
cone, and $r$ is an affine-parameter distance along each
generator. The radial coordinate $r$ also doubles as an areal radius,
meaning that $4\pi r^2$ is the area of each two-sphere 
$(v,r) = \mbox{constant}$.   

\subsection{Perturbed metric in light-cone coordinates} 

We introduce a perturbation $p_{\alpha\beta}$ of the Schwarzschild
metric, defined by the statement 
\begin{equation} 
g^{\rm perturbed}_{\alpha\beta} = g_{\alpha\beta} + p_{\alpha\beta}, 
\label{2.6}
\end{equation}
where $g^{\rm perturbed}_{\alpha\beta}$ is the metric of the perturbed
spacetime, while $g_{\alpha\beta}$ represents the Schwarzschild
solution, which we express in the light-cone coordinates of
Eq.~(\ref{2.1}). {\it We wish to place conditions on the metric
perturbation that ensure that the meaning of the light-cone
coordinates will be preserved in the perturbed spacetime.}
Specifically, we demand that in the perturbed spacetime, $v$ continues
to be constant on incoming light cones that converge toward $r=0$,
$\theta^A$ continue to be constant on the null generators of each
light cone, and $r$ continues to be an affine-parameter distance along
each generator. In exceptional circumstances that will be identified
in Sec.~II E below, $r$ also reprises its role as an areal radius,
but in general this property will not be preserved in the perturbed
spacetime.  

The geometrical meaning of the coordinates will be preserved if 
Eqs.~(\ref{2.4}) and (\ref{2.5}) continue to hold in the perturbed  
spacetime. We now have $l_\alpha = (g_{\alpha\beta} + p_{\alpha\beta})
l^\beta$, and we infer that the perturbation must satisfy the 
{\it gauge conditions} 
\begin{equation} 
p_{\alpha\beta} l^\beta = 0 \qquad \Rightarrow \qquad 
p_{vr} = p_{rr} = p_{rA} = 0. 
\label{2.7}
\end{equation} 
There are four conditions, which we refer to as the {\it light-cone
gauge conditions}. A metric perturbation $p_{\alpha\beta}$ that
satisfies Eqs.~(\ref{2.7}) will be said to be in a light-cone
gauge. As we shall see below, Eqs.~(\ref{2.7}) do not completely fix
the gauge, and the remaining gauge freedom can be exploited to
simplify the form of the perturbed metric.   

The gauge conditions leave $p_{vv}$, $p_{vA}$, and $p_{AB}$ as 
nonvanishing components of the metric perturbation. The trace of the
perturbation is
\begin{equation} 
p := g^{\alpha\beta} p_{\alpha\beta} = r^{-2} \Omega^{AB} p_{AB},  
\label{2.8}
\end{equation} 
where $\Omega^{AB}$ is the matrix inverse of $\Omega_{AB}$, defined by 
Eq.~(\ref{2.3}). The determinant of the perturbed metric is given by  
\begin{equation} 
\sqrt{-g^{\rm perturbed}} = r^2\sin\theta 
\Bigl(1 + \frac{1}{2} p \Bigr), 
\label{2.9} 
\end{equation} 
and $r$ will retain its role as areal radius whenever the metric
perturbation has a vanishing trace. In Sec.~II E we will determine
under what conditions this happens.   

\subsection{Even-parity sector} 

The even-parity sector refers to those components of the metric
perturbation that can be expanded in terms of even-parity
spherical harmonics $Y^{lm}$, $Y_A^{lm}$, $\Omega_{AB} Y^{lm}$, and 
$Y_{AB}^{lm}$. (Throughout the paper we use the notation of Martel \&
Poisson \cite{martel-poisson:05}.) The scalar harmonics $Y^{lm}$ are
the usual spherical-harmonic functions, the vectorial harmonics are
defined by $Y_A^{lm} := D_A Y^{lm}$ (where $D_A$ is the covariant
derivative operator compatible with $\Omega_{AB}$), and the tensorial
harmonics are defined by $Y_{AB}^{lm} := [D_A D_B + \frac{1}{2} l(l+1) 
\Omega_{AB}] Y^{lm}$; these are tracefree by virtue of the eigenvalue
equation for the spherical harmonics: $\Omega^{AB} Y_{AB}^{lm} = 
[\Omega^{AB} D_A D_B + l(l+1)] Y^{lm} = 0$.  

The even-parity sector is 
\begin{eqnarray} 
p_{ab} &=& \sum_{lm} h_{ab}^{lm}(x^a) Y^{lm}(\theta^A), 
\label{2.10} \\ 
p_{aB} &=& \sum_{lm} j_a^{lm}(x^a) Y_B^{lm}(\theta^A), 
\label{2.11} \\ 
p_{AB} &=& r^2 \sum_{lm} \Bigl[ K^{lm}(x^a) \Omega_{AB}
  Y^{lm}(\theta^A) 
\nonumber \\ & & \qquad \quad \mbox{} 
+ G^{lm}(x^a) Y_{AB}^{lm}(\theta^A) \Bigr], 
\label{2.12}
\end{eqnarray} 
where $x^a = (x^0,x^1) = (v,r)$. The sums over the integer $l$ begin
at $l=0$ in the case of Eq.~(\ref{2.10}) and the $K^{lm}$ term in
Eq.~(\ref{2.12}), at $l=1$ in the case of Eq.~(\ref{2.11}), and at
$l=2$ in the case of the $G^{lm}$ term in Eq.~(\ref{2.12}). The sums
over the integer $m$ go from $-l$ to $+l$. The light-cone gauge
conditions are  
\begin{equation} 
h^{lm}_{vr} = h^{lm}_{rr} = j^{lm}_r = 0. 
\label{2.13} 
\end{equation} 
The components $h^{lm}_{vv}$, $j^{lm}_v$, $K^{lm}$, and $G^{lm}$ are
nonzero in the light-cone gauge.  

Even-parity gauge transformations are generated by a dual vector field
$\Xi_\alpha = (\Xi_a,\Xi_A)$ that can be expanded as 
\begin{eqnarray} 
\Xi_a &=& \sum_{lm} \xi_a^{lm}(x^a) Y^{lm}(\theta^A), 
\label{2.14} \\ 
\Xi_A &=& \sum_{lm} \xi^{lm}(x^a) Y_A^{lm}(\theta^A). 
\label{2.15}
\end{eqnarray} 
According to Eqs.~(4.6)--(4.9) of Martel \& Poisson
\cite{martel-poisson:05}, such a transformation will preserve the
conditions of Eq.~(\ref{2.13}) provided that $\xi_v$, $\xi_r$, and
$\xi$ satisfy the equations  
\begin{eqnarray*}
0 &=& \frac{\partial \xi_v}{\partial r} 
+ \frac{\partial \xi_r}{\partial v} + \frac{2M}{r^2} \xi_r, 
\\
0 &=& \frac{\partial \xi_r}{\partial r}, 
\\
0 &=& \frac{\partial \xi}{\partial r} + \xi_r - \frac{2}{r} \xi. 
\end{eqnarray*} 
(We henceforth omit the spherical-harmonic indices for brevity. Our
considerations momentarily exclude the special cases $l=0$ and $l=1$,
which will be handled separately below.) This means that a gauge  
transformation generated by   
\begin{eqnarray} 
\xi_v &=& -\dot{a}(v) r - f a(v) + b(v), 
\label{2.16} \\ 
\xi_r &=& a(v), 
\label{2.17} \\
\xi &=& a(v) r + c(v) r^2 
\label{2.18}
\end{eqnarray}
will keep a perturbation within the light-cone gauge. The remaining 
gauge freedom is therefore characterized by three arbitrary functions  
$a(v)$, $b(v)$, $c(v)$, and the overdot in Eq.~(\ref{2.16}) indicates
differentiation with respect to $v$. The gauge transformation changes
the nonvanishing components of the perturbation field according to   
\begin{eqnarray}
h_{vv} \to h'_{vv} &=& h_{vv} + 2 \ddot{a}\, r + 2 \biggl( 1 -
\frac{3M}{r} \biggr) \dot{a} - 2 \dot{b} 
\nonumber \\ & & \mbox{} 
+ \frac{2M}{r^2} b, 
\label{2.19} \\
j_v \to j'_v &=& j_v + f a - b - \dot{c}\, r^2, 
\label{2.20} \\ 
K \to K' &=& K + 2\dot{a} + \frac{l(l+1)}{r} a - \frac{2}{r} b 
\nonumber \\ & & \mbox{} 
+ l(l+1)c, 
\label{2.21} \\ 
G \to G' &=& G - \frac{2}{r} a - 2c. 
\label{2.22}
\end{eqnarray} 

The lower multipoles $l=0$ and $l=1$ must be considered
separately. For $l=0$ the spherical harmonics $Y_A$ and $Y_{AB}$ are 
identically zero, and the only relevant perturbation fields are
$h_{ab}$ and $K$; $j_a$ and $G$ are not defined. A gauge
transformation is then generated by $\Xi_a = \xi_a Y^{00}$, $\Xi_A  
= 0$, and $\xi$ is not defined. It is easy to check that the
light-cone gauge will be preserved with a $\xi_a$ that is still given
by Eqs.~(\ref{2.16}) and (\ref{2.17}). In this case the remaining
gauge freedom is characterized by two arbitrary functions, $a(v)$ and 
$b(v)$. The corresponding change in $h_{vv}$ is still given by
Eq.~(\ref{2.19}), while $K'$ is obtained by setting $l=0$ in
Eq.~(\ref{2.21}). For $l=1$ the tensorial harmonics $Y_{AB}$ are
identically zero, and only $G$ is not defined. The gauge
transformation of Eqs.~(\ref{2.16})--(\ref{2.18}) is still seen to
preserve the light-cone gauge, and the changes in the perturbation
fields are still described by Eqs.~(\ref{2.19})--(\ref{2.21}), in
which we must set $l(l+1) = 2$; Eq.~(\ref{2.22}) is irrelevant when 
$l = 1$.      

It is easy to verify that the dual vector field of
Eqs.~(\ref{2.14})--(\ref{2.18}) generates the (small) coordinate
transformation 
\begin{eqnarray}
v &\to& v' = v + a(v,\theta^A), 
\label{2.23} \\ 
r &\to& r' = \biggl(1 - \frac{\partial a}{\partial v} \biggr) r 
+ b(v,\theta^A), 
\label{2.24} \\ 
\theta^A &\to& \theta^{\prime A} = \theta^A + \Omega^{AB} 
\frac{\partial}{\partial \theta^B} \biggl[ \frac{a}{r} + c(v,\theta^A)
  \biggr], 
\label{2.25}
\end{eqnarray} 
where $a(v,\theta^A) := \sum_{lm} a^{lm}(v) Y^{lm}(\theta^A)$, with
similar equations defining $b(v,\theta^A)$ and $c(v,\theta^A)$. 
This transformation leaves the conditions of Eq.~(\ref{2.7}) intact.    

\subsection{Odd-parity sector} 

The odd-parity sector refers to those components of the metric
perturbation that can be expanded in terms of odd-parity spherical 
harmonics $X_A^{lm}$ and $X_{AB}^{lm}$. The vectorial harmonics are
defined by $X_A^{lm} := -\varepsilon_A^{\ B} D_B Y^{lm}$, where 
$\varepsilon_{AB}$ is the Levi-Civita tensor on the unit two-sphere
(with independent component $\varepsilon_{\theta\phi} = \sin\theta$),
and where $\varepsilon_A^{\ B} := \Omega^{BC} \varepsilon_{AC}$. The
tensorial harmonics are $X_{AB}^{lm} := -\frac{1}{2} 
(\varepsilon_A^{\ C} D_B + \varepsilon_B^{\ C} D_A) D_C Y^{lm}$; these 
are tracefree by virtue of the antisymmetry of the Levi-Civita tensor: 
$\Omega^{AB} X_{AB}^{lm} = -\varepsilon^{AB} D_A D_B Y^{lm} = 0$.  

The odd-parity sector is 
\begin{eqnarray} 
p_{ab} &=& 0, 
\label{2.26} \\ 
p_{aB} &=& \sum_{lm} h_a^{lm}(x^a) X_B^{lm}(\theta^A), 
\label{2.27} \\ 
p_{AB} &=& \sum_{lm} h_2^{lm}(x^a) X_{AB}^{lm}(\theta^A). 
\label{2.28}
\end{eqnarray} 
The sums over the integer $l$ begin at $l=1$ in the case of
Eq.~(\ref{2.27}), and at $l=2$ in the case of Eq.~(\ref{2.28}). The
light-cone gauge conditions are    
\begin{equation} 
h^{lm}_r = 0. 
\label{2.29} 
\end{equation} 
The components $h^{lm}_v$ and $h^{lm}_2$ are nonzero in the
light-cone gauge.  

Odd-parity gauge transformations are generated by a dual vector field 
$\Xi_\alpha = (\Xi_a,\Xi_A)$ that can be expanded as 
\begin{eqnarray} 
\Xi_a &=& 0, 
\label{2.30} \\ 
\Xi_A &=& \sum_{lm} \xi^{lm}(x^a) X_A^{lm}(\theta^A). 
\label{2.31}
\end{eqnarray} 
According to Eqs.~(5.5) of Martel \& Poisson \cite{martel-poisson:05},
such a transformation will preserve the conditions of Eq.~(\ref{2.29})
provided that $\xi$ satisfies $\partial \xi/\partial r - 2\xi/r 
= 0$. (We henceforth omit the spherical-harmonic indices for
brevity. Our considerations momentarily exclude the special case
$l=1$, which will be handled separately below.) This means that a
gauge transformation generated by     
\begin{equation} 
\xi = \alpha(v) r^2 
\label{2.32}
\end{equation} 
will keep a perturbation within the light-cone gauge. The remaining 
gauge freedom is therefore characterized by a single arbitrary
function $\alpha(v)$. The gauge transformation changes the
nonvanishing components of the perturbation field according to   
\begin{eqnarray}
h_{v} &\to& h'_{v} = h_{v} - \dot{\alpha}\, r^2, 
\label{2.33} \\
h_2 &\to& h'_2 = h_2 - 2 \alpha\, r^2, 
\label{2.34}
\end{eqnarray}
where an overdot indicates differentiation with respect to $v$. 

The situation is the same for the special case $l=1$, except that  
$X_{AB}$ is then identically zero and $h_2$ is not defined. The gauge 
transformation of Eq.~(\ref{2.32}) is still seen to preserve the
light-cone gauge, and it still changes $h_v$ according to
Eq.~(\ref{2.33}); Eq.~(\ref{2.34}) is then irrelevant.    

It is easy to verify that the dual vector field of
Eqs.~(\ref{2.30})--(\ref{2.32}) generates the (small) coordinate   
transformation
\begin{equation}
\theta^A \to \theta^{\prime A} = \theta^A - \varepsilon^{AB}  
\frac{\partial}{\partial \theta^B} \alpha(v,\theta^A), 
\label{2.35}
\end{equation} 
where $\alpha(v,\theta^A) := \sum_{lm} \alpha^{lm}(v)
Y^{lm}(\theta^A)$ and $\varepsilon^{AB} := \Omega^{AC} \Omega^{BD}
\varepsilon_{CD}$. This transformation leaves the conditions of 
Eq.~(\ref{2.7}) intact. 

\subsection{When is $r$ an areal radius?} 

According to Eqs.~(\ref{2.8}) and (\ref{2.9}), $r$ keeps its
interpretation as an areal radius whenever $\Omega^{AB} p_{AB} 
= 0$. And according to Eq.~(\ref{2.12}), this happens when
$K^{lm}(v,r) = 0$ for all values of $l$ and $m$. In this subsection we
determine under what circumstances it is possible to impose this
condition.     

The light-cone gauge produces a very convenient decoupling of the
equation that governs the behavior of $K^{lm}$ from the equations that  
determine the remaining perturbation fields. According to the
field equations listed in the Appendix, we have 
\begin{equation} 
Q^{vv}_{lm} = Q^{lm}_{rr} = -\frac{\partial^2}{\partial r^2} K^{lm}  
- \frac{2}{r} \frac{\partial}{\partial r} K^{lm}, 
\label{2.36}
\end{equation}
where, for example, $Q^{lm}_{rr} := 8\pi \int T_{rr} \bar{Y}^{lm}\,
d\Omega$, with $d\Omega = \sin\theta\, d\theta d\phi$, are the
spherical-harmonic projections of the $rr$ component of the
energy-momentum tensor. When $T^{vv} = T_{rr} = 0$, Eq.~(\ref{2.36})
reveals that $K^{lm}$ must be of the form $p^{lm}(v) + q^{lm}(v)/r$,
where $p^{lm}$ and $q^{lm}$ are arbitrary functions of $v$. But it is
possible to exploit the remaining gauge freedom contained in
Eqs.~(\ref{2.16})--(\ref{2.18}) to set $K^{\prime lm} = 0$. As can be
seen from Eq.~(\ref{2.21}), this condition constrains the functions 
$b^{lm}(v)$ and $c^{lm}(v)$, which must now be related to
$a^{lm}(v)$. The remaining gauge freedom is therefore restricted to
transformations characterized by a single arbitrary function,
$a^{lm}(v)$. Our conclusion is that $K^{lm}$ can be set equal to zero
whenever $T^{vv} = T_{rr} = 0$, and that this operation still does not
fully exhaust the gauge freedom.     

We have established the following theorem: When the energy-momentum  
tensor responsible for the metric perturbation is such that 
\begin{equation} 
T_{\alpha\beta} l^\alpha l^\beta = 0,  
\label{2.37} 
\end{equation} 
the light-cone gauge can be refined to include the tracefree condition   
\begin{equation} 
p := g^{\alpha\beta} p_{\alpha\beta} = 0
\label{2.38}
\end{equation}  
in addition to the four conditions of Eq.~(\ref{2.7}). In such 
circumstances, $\sqrt{-g^{\rm perturbed}} = r^2\sin\theta$ and $r$ 
retains its interpretation as an areal radius. In these circumstances
the light-cone gauge becomes the ``incoming radiation gauge'' of
Chrzanowski \cite{chrzanowski:75a, wald:78, lousto-whiting:02,
ori:03}.     

The theorem relating the tracefree condition of Eq.~(\ref{2.38}) to
the vanishing of $T_{\alpha\beta} l^\alpha l^\beta$ is a new
result. The theorem was established independently by Price, Shankar,
and Whiting \cite{price:05, price-etal:06} in work that has not yet
been published, except for a statement of the result made in Sec.~4.1
of Ref.~\cite{whiting-price:05}. Remarkably, these authors were able 
to extend the theorem from Schwarzschild spacetime to all Petrov
type-II spacetimes.     

\subsection{Discussion; Comparison with the Regge-Wheeler gauge}  

The light-cone gauge possesses two main virtues. The first is that 
it involves simple algebraic conditions on the metric perturbation;
these were stated in covariant form in Eq.~(\ref{2.7}),
$p_{\alpha\beta} l^\beta = 0$, and they were stated in expanded form
in Eqs.~(\ref{2.13}) and (\ref{2.29}), $h^{lm}_{vr} = h^{lm}_{rr} =
j^{lm}_r = h^{lm}_r = 0$. The second is that the gauge conditions
preserve the geometrical meaning of the original coordinate system
$(v,r,\theta^A)$; as was shown in Sec.~II B, the advanced-time
coordinate $v$ continues to label incoming light cones that converge 
toward $r=0$, the angles $\theta^A$ continue to label the generators
of each light cone, and the radial coordinate $r$ continues to be an
affine-parameter distance along each generator. The task of extracting 
the physical properties of a perturbed spacetime will be greatly
facilitated by the use of such meaningful coordinates.  

Most of the literature on black-hole perturbation theory employs an
alternative gauge known as the ``Regge-Wheeler gauge''
\cite{regge-wheeler:57}. The gauge conditions in this case are
$j^{lm}_v = j^{lm}_r = G^{lm} = h^{lm}_2 = 0$. The Regge-Wheeler gauge
also has the advantage of involving simple algebraic conditions on the
metric perturbation. But unlike the light-cone gauge, the
Regge-Wheeler gauge produces a coordinate system that does not possess
a clear geometrical meaning; this is a disadvantage. And indeed,
the coordinates can sometimes be pathological. For example, the
Regge-Wheeler gauge produces metric components that do not display
asymptotically-flat behavior near future null infinity, even when the
source of the perturbation is spatially bounded
\cite{gleiser-etal:00}. This problem is associated with the fact that
by imposing $G^{lm} = h^{lm}_2 = 0$, the Regge-Wheeler gauge is
actually setting to zero the transverse-tracefree part of the metric
perturbation, thereby effectively ``gauging away'' its
gravitational-wave content. (The gravitational-wave modes are still  
present in the metric perturbation, but in the Regge-Wheeler gauge
they are encoded in unnatural places.) The end result is a meaningless  
coordinate system, a metric perturbation that fails to be
asymptotically flat, and a spacetime that does not easily reveal its
radiation content. These problems are not present in the light-cone
gauge.  

Another approach that has been followed in the literature on 
black-hole perturbation theory is to avoid fixing the gauge, and to
work instead with a gauge-invariant formalism \cite{moncrief:74, 
gerlach-sengupta:79, sarbach-tiglio:01, gundlach-martingarcia:00,  
clarkson-barrett:03, nagar-rezzolla:05, martel-poisson:05}. Such an
approach can be very useful, especially when an application calls for
a switch from one gauge to another. We would argue, however, that 
{\it a good choice of gauge} can be even more useful in concrete
applications. After all, most relativists would begin an investigation
of the Schwarzschild spacetime by making a specific choice of
coordinate system; few relativists would insist on staying coordinate
free. And most relativists would agree that the Eddington-Finkelstein
system $(v,r,\theta^A)$ is more convenient to work with than the
Schwarzschild coordinates $(t,r,\theta^A)$ when one is concerned with
the event horizon; these relativists would say that the
Eddington-Finkelstein coordinates are {\it good coordinates}. These  
attitudes need not change when one goes slightly away from the
Schwarzschild spacetime, and the light-cone gauge provides a 
{\it good coordinate system} to investigate the perturbed spacetimes.  

\section{Black hole in a magnetic field}           
 
\subsection{Physical situation} 

To illustrate the use of the light-cone gauge in black-hole
perturbation theory, we work through a model problem involving a black 
hole immersed in a uniform magnetic field. We have in mind a situation
in which a large mechanical structure, such as a giant solenoid, is
set up in outer space and made to produce a uniform magnetic field of 
strength $B$. The structure has a mass $M'$ and its linear extension
is of the order of the length scale\footnote{The constant $a$ is not
to be confused with the functions $a^{lm}(v)$ introduced in Sec.~II
C.} $a$; the magnetic field is imagined to be uniform over a region of
this size. A black hole of mass $M$, initially isolated, is then
brought to the structure and inserted within the magnetic field. This
process is quasi-static and reversible, and the black hole's surface
area stays constant during the immersion. We wish to study how the
black hole distorts the magnetic field within the structure, and how
the magnetic field distorts the geometry of the black hole.  

We suppose that the perturbation created by the magnetic field is 
small and that its effects can be adequately calculated with the
equations of black-hole perturbation theory. We shall see below that
the criterion for this is $r^2 B^2 \ll 1$, where $r$ is the distance
from the black hole. If we restrict our attention to the interior of
the mechanical structure and impose the inequality $r < a$, then the
perturbative criterion is  
\begin{equation} 
a^2 B^2 \ll 1. 
\label{3.1}
\end{equation} 
In addition to Eq.~(\ref{3.1}) we assume that the structure is
situated in the black hole's weak-field region, so that 
\begin{equation} 
\frac{M}{a} \ll 1. 
\label{3.2}
\end{equation} 
While $a^2 B^2$ and $M/a$ must both be small, their relative sizes are
not constrained. We may imagine that $M/a$ is either much smaller
than, comparable to, or much larger than $a^2 B^2$; black-hole
perturbation theory can handle all these situations. Below we will be 
particularly (but not exclusively) interested in the first
possibility, $M/a \ll a^2 B^2$ or $M/a^3 \ll B^2$. In this situation 
there exists an asymptotic region (described by $M \ll r < a$) in
which the gravitational effects of the magnetic field, though small,
are larger than those of the black hole.     

Another aspect of our model problem is the tidal gravity exerted by
the mechanical structure. Because the structure has a mass $M'$ and is
situated at a distance $a$ from the black hole, the tidal field (or
Weyl curvature) it produces near the black hole is ${\cal E} \sim
M'/a^3$. This quantity ${\cal E}$, which will be formally introduced
below, is an additional parameter that characterizes the physical
situation. Below we will imagine that ${\cal E}$ is of the same order
of magnitude as $B^2$, so that $M'/a^3 \sim B^2$. Our results,
however, will not be tied by this assumption; they will be just as
valid when ${\cal E}$ is much smaller than (or indeed much larger
than) $B^2$.       

The perturbed black-hole solution that we construct below is in fact a 
three-parameter family of solutions; each solution is characterized by 
the black-hole mass $M$, the magnetic field strength $B$, and the
tidal gravity ${\cal E}$. The solution is obtained perturbatively
through order $(B^2, {\cal E})$. 

\subsection{Magnetic field} 

We first calculate the electromagnetic field $F_{\alpha\beta}$ that
surrounds the black hole. Because we seek to determine the perturbed
metric accurately through order $B^2$, it is sufficient to calculate
$F_{\alpha\beta}$ to order $B$. And because the metric corrections of 
order $B^2$ do not enter this calculation, we may let the spacetime
have an unperturbed Schwarzschild metric.   

To find the electromagnetic field we rely on Wald's observation
\cite{wald:74} that in a vacuum spacetime, any Killing vector can be
identified with the vector potential of a test electromagnetic field. 
The fact that the vector satisfies Killing's equation ensures that the
resulting $F_{\alpha\beta}$ satisfies the sourcefree Maxwell
equations. To produce a magnetic field that is asymptotically uniform
when $r \gg M$, we set     
\begin{equation} 
A^\alpha = \frac{1}{2} B \phi^\alpha, 
\label{3.3}
\end{equation}
where $\phi^\alpha := (0,0,0,1)$ is the rotational Killing vector of
the unperturbed Schwarzschild spacetime; we use the ordering
$(v,r,\theta,\phi)$ of the unperturbed light-cone coordinates. 

The vector potential gives rise to an electromagnetic field tensor 
$F_{\alpha\beta} = \nabla_\alpha A_\beta - \nabla_\beta A_\alpha$. To
display its components it is useful to decompose it in an orthonormal
tetrad $e^\alpha_\mu$ that is oriented along the ``Cartesian
directions'' associated with the ``spherical coordinates''
$(r,\theta,\phi)$; here the superscript $\alpha$ is the usual
vectorial index, and the subscript $\mu$ is a frame index that 
identifies each member of the tetrad. We thus introduce the tetrad   
\begin{eqnarray}
e^\alpha_0 &=& \bigl( f^{-1/2},0,0,0 \bigr),
\label{3.4} \\ 
e^\alpha_1 &=& \bigl( f^{-1/2} \sin\theta\cos\phi,
f^{1/2}\sin\theta\cos\phi, r^{-1} \cos\theta\cos\phi, 
\nonumber \\ & & 
-r^{-1} \sin\phi/\sin\theta \bigr), 
\label{3.5} \\ 
e^\alpha_2 &=& \bigl( f^{-1/2} \sin\theta\sin\phi,
f^{1/2}\sin\theta\sin\phi, r^{-1} \cos\theta\sin\phi, 
\nonumber \\ & & 
r^{-1} \cos\phi/\sin\theta \bigr), 
\label{3.6} \\ 
e^\alpha_3 &=& \bigl( f^{-1/2}\cos\theta, f^{1/2}\cos\theta,  
-r^{-1}\sin\theta, 0 \bigr). 
\label{3.7}
\end{eqnarray} 
We may think of $e^\alpha_1$ as pointing in the ``$x$-direction,'' of 
$e^\alpha_2$ as pointing in the ``$y$-direction,'' and of $e^\alpha_3$
as pointing in the ``$z$-direction.'' In this tetrad, the nonvanishing 
frame components of the electromagnetic field tensor are 
\begin{eqnarray} 
B_1 &:=& F_{23} := F_{\alpha\beta} e^\alpha_2 e^\beta_3 
\nonumber \\ 
&=& B \bigl( 1 - \sqrt{f} \bigr) \sin\theta\cos\theta\cos\phi, 
\label{3.8} \\ 
B_2 &:=& F_{31} := F_{\alpha\beta} e^\alpha_3 e^\beta_1 
\nonumber \\ 
&=& B \bigl( 1 - \sqrt{f} \bigr) \sin\theta\cos\theta\sin\phi, 
\label{3.9} \\ 
B_3 &:=& F_{12} := F_{\alpha\beta} e^\alpha_1 e^\beta_2 
\nonumber \\ 
&=& B \bigl[ \sqrt{f} + \bigl( 1 - \sqrt{f} \bigr) \cos^2\theta \bigr],   
\label{3.10}
\end{eqnarray} 
where $f := 1-2M/r$. The field is purely magnetic. Asymptotically,
when $r \gg M$, $B_1 \sim 0$, $B_2 \sim 0$, $B_3 \sim B$; the
magnetic field is uniform and aligned with the $z$-direction. Closer
to the black hole the magnetic field is distorted; at $r=2M$ we have  
$|\bm{B}|^2 := B_1^2 + B_2^2 + B_3^3 = B^2 \cos^2\theta$, which
indicates that the field is strongest at the poles. 

The electromagnetic field produces an energy-momentum tensor given by 
\begin{equation}
T^{\alpha\beta} = \frac{1}{4\pi} \biggl( F^{\alpha\gamma} 
F^\beta_{\ \gamma} - \frac{1}{4} g^{\alpha\beta} F^{\gamma\delta}
F_{\gamma\delta} \biggr). 
\label{3.11}
\end{equation} 
Its nonvanishing components are 
\begin{eqnarray} 
\hspace*{-15pt}
T^{vv} &=& \frac{B^2}{4\pi} \sin^2\theta, 
\label{3.12} \\ 
\hspace*{-15pt}
T^{vr} &=& \frac{B^2}{8\pi} \frac{1}{r} \bigl[ r - 2M 
    - 2(r-M)\cos^2\theta \bigr], 
\label{3.13} \\ 
\hspace*{-15pt}
T^{rr} &=& \frac{B^2}{8\pi} \frac{r-2M}{r^2} \bigl[ r - 2M 
    - 2(r-M)\cos^2\theta \bigr],
\label{3.14} \\ 
\hspace*{-15pt}
T^{v\theta} &=& \frac{B^2}{4\pi} \frac{1}{r} \sin\theta\cos\theta, 
\label{3.15} \\ 
\hspace*{-15pt}
T^{r\theta} &=& \frac{B^2}{4\pi} \frac{r-2M}{r^2}
\sin\theta\cos\theta,  
\label{3.16} \\ 
\hspace*{-15pt}
T^{\theta\theta} &=& -\frac{B^2}{8\pi} \frac{1}{r^3} \bigl[ r - 2M  
    - 2(r-M)\cos^2\theta \bigr], 
\label{3.17} \\ 
\hspace*{-15pt}
T^{\phi\phi} &=& \frac{B^2}{8\pi} \frac{1}{r^3\sin^2\theta} 
\bigl[ r - 2M + 2M\cos^2\theta \bigr].  
\label{3.18} 
\end{eqnarray} 
This energy-momentum tensor is the source of the metric perturbation
that will be calculated in the following subsections. 

It is easy to see from Eqs.~(\ref{3.12})--(\ref{3.18}) that the
angular dependence of the energy-momentum tensor is contained entirely
in spherical-harmonic functions of degrees $l=0$ and $l=2$; and
because there is no dependence on $\phi$, only functions with
azimuthal index $m=0$ are involved. It can also be seen that the
angular dependence of the energy-momentum tensor has an even
parity. Our solution to the equations of black-hole perturbation
theory will therefore have the following properties: (i) it will be
axially symmetric; (ii) it will contain even-parity spherical-harmonic
modes with $(l,m) = \{ (0,0), (2,0) \}$ only; and (iii) it will be
stationary. The metric perturbation will contain a term of magnetic
origin, and it will also contain a homogeneous term associated with
the ambient Weyl curvature.          

\subsection{Integrating the perturbation equation: $l=0$} 

As discussed in Sec.~IV D of Martel \& Poisson
\cite{martel-poisson:05} (see also the Appendix of this paper), the
relevant projections of the energy-momentum tensor when $l=0$ are
$Q^{ab}$ and $Q^\flat$, which are defined in the Appendix. Using
$Y^{00} = 1/\sqrt{4\pi}$ and the energy-momentum tensor of
Eqs.~(\ref{3.12})--(\ref{3.18}), we obtain  
\begin{eqnarray} 
Q^{vv} &=& 4 b^2, 
\label{3.19} \\ 
Q^{vr} &=& b^2 \frac{r-4M}{r}, 
\label{3.20} \\ 
Q^{rr} &=& b^2 \frac{(r-2M)(r-4M)}{r^2}, 
\label{3.21} \\ 
Q^\flat &=& 2 b^2, 
\label{3.22}
\end{eqnarray} 
where\footnote{The constant $b$ is not to be confused with the
functions $b^{lm}(v)$ introduced in Sec.~II C.} 
$b^2 := 2\sqrt{\pi} B^2/3$.  

We now integrate the perturbation equations for the two relevant
functions $K(r)$ and $h_{vv}(r)$ --- please refer to the listing of
field equations in the Appendix. We first substitute Eq.~(\ref{3.19})
into Eq.~(\ref{2.36}) and solve for $K$. The general solution is $K(r) 
= -\frac{2}{3} b^2 r^2 + p + q/r$, where $p$ and $q$ are arbitrary
constants. As discussed in Secs.~II C and II E, we may exploit the
remaining gauge freedom to set them equal to zero. We have, therefore,    
\begin{equation} 
K = -\frac{2}{3} b^2 r^2. 
\label{3.23}
\end{equation} 
The remaining field equations provide a number of equivalent
differential equations for $h_{vv}$. The general solution is
$h_{vv}(r) = -\frac{1}{3} b^2 r(3r - 8M) + 2\delta M/r$. It involves
an arbitrary constant $\delta M$ that can be interpreted as a shift in
$M$, the black-hole mass parameter. To reflect the fact that we wish
our perturbed black hole to have the same surface area as our
original, unperturbed black hole (this was motivated back in Sec.~III
A), we set $\delta M = 0$. We will verify in Sec.~III F that this
condition does indeed lead to a preservation of the horizon area. We
have, therefore, 
\begin{equation} 
h_{vv} = -\frac{1}{3} b^2 r(3r - 8M). 
\label{3.24} 
\end{equation} 

Substituting Eqs.~(\ref{3.23}) and (\ref{3.24}) into
Eqs.~(\ref{2.10})--(\ref{2.12}) yields
\begin{eqnarray} 
p_{vv} &=& -\frac{1}{9} B^2 r (3r - 8M), 
\label{3.25} \\ 
p_{\theta\theta} &=& -\frac{2}{9} B^2 r^4, 
\label{3.26} \\
p_{\phi\phi} &=& -\frac{2}{9} B^2 r^4 \sin^2\theta 
\label{3.27} 
\end{eqnarray} 
for the $l=0$ sector of the metric perturbation. 

\subsection{Integrating the perturbation equation: $l=2$} 

The relevant spherical-harmonic functions are $Y^{20} = \frac{1}{4}
\sqrt{5/\pi} (3\cos^2\theta - 1)$, $Y^{20}_\theta = -\frac{3}{2} 
\sqrt{5/\pi} \sin\theta\cos\theta$, $Y^{20}_\phi = 0$,
$Y^{20}_{\theta\theta} = \frac{3}{4} \sqrt{5/\pi} \sin^2\theta$,
$Y^{20}_{\theta\phi} = 0$, and $Y^{20}_{\phi\phi} = -\frac{3}{4}
\sqrt{5/\pi} \sin^4\theta$. The required projections of the
energy-momentum tensor are defined in Eqs.~(4.17)--(4.20) of 
Martel \& Poisson \cite{martel-poisson:05} (see also the Appendix of
this paper); they are   
\begin{eqnarray} 
Q^{vv} &=& -b^2, 
\label{3.28} \\ 
Q^{vr} &=& -b^2 \frac{r-M}{r}, 
\label{3.29} \\ 
Q^{rr} &=& -b^2 \frac{(r-2M)(r-M)}{r^2}, 
\label{3.30} \\ 
Q^v &=& -b^2 r, 
\label{3.31} \\ 
Q^r &=& -b^2 (r-2M), 
\label{3.32} \\ 
Q^\flat &=& b^2, 
\label{3.33} \\ 
Q^\sharp &=& -b^2 r(r-2M), 
\label{3.34}
\end{eqnarray}
where the constant\footnote{The constant $b$ is not to be confused
with the functions $b^{lm}(v)$ introduced in Sec.~II C.} $b^2$ has
been reassigned to $b^2 := \frac{8}{3} \sqrt{\pi/5} B^2$.  

We now integrate the field equations for the four functions
$h_{vv}(r)$, $j_v(r)$, $K(r)$, and $G(r)$. The equation for $K$
decouples, as was shown in Eq.~(\ref{2.36}), and it involves the
source term $Q^{vv}$. Exploiting the remaining gauge freedom to set
all integration constants to zero, we take the solution to be  
\begin{equation} 
K = \frac{1}{6} b^2 r^2. 
\label{3.35} 
\end{equation} 
The remaining field equations form a set of coupled ordinary
differential equations for the remaining quantities $h_{vv}$, $j_v$,
and $G$. These equations are easily decoupled by taking additional
derivatives, and we easily obtain general solutions to the
higher-order equations. These would-be solutions involve a number of  
integration constants that are not part of the true solution space;
these are determined by substituting the would-be solutions into the
original system of second-order equations, and making sure that the
solutions stay valid. At the end of this process we obtain 
$h_{vv} = -c_1 M/r^2 - 3c_2 (r-2M)^2 + \frac{1}{3} b^2 M(r-3M)$, 
$j_v = \frac{1}{3} c_1 (r+M)/r - c_2 r^2 (r-2M) 
+ \frac{1}{6} b^2 M r^2$, and 
$G = \frac{1}{3} c_1/r - c_2 (r^2 - 2M^2) 
+ \frac{1}{6} b^2 (r^2 + M^2)$, where $c_1$ and $c_2$ are the
remaining constants of integration. As we shall show below, the gauge
freedom that is still at our disposal can be exploited to set $c_1 =
0$. Setting also $c_2 = \frac{1}{6} b^2 - \frac{1}{3} \varepsilon$ for
later convenience (thus discarding $c_2$ in favor of the new constant
$\varepsilon$), our solutions are  
\begin{eqnarray} 
h_{vv} &=& -\frac{1}{6} b^2 (3r^2 - 14Mr + 18M^2) 
\nonumber \\ & & \mbox{} 
+ \varepsilon (r-2M)^2, 
\label{3.36} \\ 
j_v &=& -\frac{1}{6} b^2 r^2(r-3M) 
+ \frac{1}{3} \varepsilon r^2(r-2M), 
\label{3.37} \\ 
G &=& \frac{1}{2} b^2 M^2 + \frac{1}{3} \varepsilon (r^2 - 2M^2). 
\label{3.38} 
\end{eqnarray} 

Substituting Eqs.~(\ref{3.35})--(\ref{3.38}) into
Eqs.~(\ref{2.10})--(\ref{2.12}) yields 
\begin{eqnarray} 
p_{vv} &=& -\frac{1}{9} B^2 (3r^2 - 14Mr + 18M^2)(3\cos^2\theta - 1)   
\nonumber \\ & & \mbox{} 
+ {\cal E} (r-2M)^2 (3\cos^2\theta - 1), 
\label{3.39} \\ 
p_{v\theta} &=& \frac{2}{3} B^2 r^2(r-3M)\sin\theta\cos\theta  
\nonumber \\ & & \mbox{} 
- 2 {\cal E} r^2(r-2M) \sin\theta\cos\theta, 
\label{3.40} \\ 
p_{\theta\theta} &=& \frac{1}{9} B^2 r^4 (3\cos^2\theta - 1) 
+ B^2 M^2 r^2\sin^2\theta 
\nonumber \\ & & \mbox{} 
+ {\cal E} r^2(r^2 - 2M^2)\sin^2\theta, 
\label{3.41} \\ 
p_{\phi\phi} &=& \frac{1}{9} B^2 r^4 \sin^2\theta (3\cos^2\theta - 1)  
- B^2 M^2 r^2\sin^4\theta 
\nonumber \\ & & \mbox{} 
- {\cal E} r^2(r^2 - 2M^2)\sin^4\theta
\label{3.42}
\end{eqnarray} 
for the $l=2$ sector of the metric perturbation. We have introduced
the constant ${\cal E} := \frac{1}{4} \sqrt{5/\pi}\, \varepsilon$; its 
interpretation as a tidal gravitational field will be examined 
below. 

We must now explain why it was admissible to set $c_1 = 0$ in our 
solutions. We go back to Eqs.~(\ref{2.19})--(\ref{2.22}) and consider 
the subclass of gauge transformations that leave $K$ unchanged (in
addition to $h_{vr}$, $h_{rr}$, and $j_r$, which are all zero in the
light-cone gauge). We see that when $l=2$, the subclass is
characterized by a single function $a(v)$, with the other functions
related to it by $b(v) = 3a$ and $c(v) = -\frac{1}{3} \dot{a}$. Taking
$a$ to be a constant produces $c=0$, and we observe that under such a
gauge transformation, $h_{vv}$ changes by a term $6aM/r^2$, $j_v$
changes by a term $-2a(r+M)/r$, and $G$ changes by a term $-2a/r$. We
then see that selecting $a = \frac{1}{6} c_1$ produces a gauge
transformation that effectively sets $c_1$ to zero. There is therefore
no loss of generality in making this assignment.    

\subsection{Perturbed metric}

Combining Eqs.~(\ref{3.25})--(\ref{3.27}) from Sec.~III C and
Eqs.~(\ref{3.39})--(\ref{3.42}) from Sec.~III D gives us the metric of
our perturbed black hole. Its nonvanishing components are  
\begin{eqnarray} 
g_{vv} &=& -f - \frac{1}{9} B^2 r (3r - 8M) 
\nonumber \\ & & \mbox{} 
- \frac{1}{9} B^2 (3r^2 - 14Mr + 18M^2)(3\cos^2\theta - 1) 
\nonumber \\ & & \mbox{} 
+ {\cal E} (r-2M)^2 (3\cos^2\theta - 1), 
\label{3.43} \\ 
g_{vr} &=& 1, 
\label{3.44} \\ 
g_{v\theta} &=& \frac{2}{3} B^2 r^2(r-3M)\sin\theta\cos\theta  
\nonumber \\ & & \mbox{} 
- 2 {\cal E} r^2(r-2M) \sin\theta\cos\theta,
\label{3.45} \\ 
g_{\theta\theta} &=& r^2 - \frac{2}{9} B^2 r^4 
+ \frac{1}{9} B^2 r^4 (3\cos^2\theta - 1) 
\nonumber \\ & & \mbox{} 
+ B^2 M^2 r^2\sin^2\theta 
\nonumber \\ & & \mbox{} 
+ {\cal E} r^2(r^2 - 2M^2)\sin^2\theta, 
\label{3.46} \\ 
g_{\phi\phi} &=& r^2\sin^2\theta - \frac{2}{9} B^2 r^4\sin^2\theta 
\nonumber \\ & & \mbox{} 
+ \frac{1}{9} B^2 r^4 \sin^2\theta (3\cos^2\theta - 1) 
\nonumber \\ & & \mbox{} 
- B^2 M^2 r^2\sin^4\theta 
\nonumber \\ & & \mbox{} 
- {\cal E} r^2(r^2 - 2M^2)\sin^4\theta, 
\label{3.47}
\end{eqnarray} 
and we observe that the perturbation is small whenever $r^2 B^2 \ll
1$ and $r^2 {\cal E} \ll 1$, as was anticipated in Sec.~III A. This is
a three-parameter family of solutions to the Einstein-Maxwell
equations, accurate through order $(B^2, {\cal E})$. The
electromagnetic field is generated by the vector potential of
Eq.~(\ref{3.3}); it is accurate through order $B$. The parameters of
the family are the black-hole mass $M$, the magnetic field strength
$B$, and the tidal gravitational field ${\cal E}$.   

The interpretation of ${\cal E}$ as a tidal-gravity (Weyl-curvature)
parameter comes from an examination of the asymptotic behavior of the
metric when $r \gg M$ (keeping $r \ll 1/B$, as was discussed in
Sec.~III A). In this regime Eqs.~(\ref{3.43})--(\ref{3.47}) reduce to   
\begin{eqnarray}  
g_{vv} &\sim& -1 - \frac{1}{3} B^2 r^2 
- \frac{1}{3} B^2 r^2 (3\cos^2\theta - 1) 
\nonumber \\ & & \mbox{} 
+ {\cal E} r^2 (3\cos^2\theta - 1), 
\label{3.48} \\ 
g_{vr} &=& 1, 
\label{3.49} \\ 
g_{v\theta} &\sim& \frac{2}{3} B^2 r^3 \sin\theta\cos\theta  
- 2 {\cal E} r^3 \sin\theta\cos\theta, 
\label{3.50} \\ 
g_{\theta\theta} &\sim& r^2 - \frac{2}{9} B^2 r^4 
+ \frac{1}{9} B^2 r^4 (3\cos^2\theta - 1) 
\nonumber \\ & & \mbox{} 
+ {\cal E} r^4 \sin^2\theta, 
\label{3.51} \\ 
g_{\phi\phi} &\sim& r^2\sin^2\theta - \frac{2}{9} B^2 r^4\sin^2\theta  
\nonumber \\ & & \mbox{} 
+ \frac{1}{9} B^2 r^4 \sin^2\theta (3\cos^2\theta - 1) 
\nonumber \\ & & \mbox{} 
- {\cal E} r^4 \sin^4\theta.
\label{3.52} 
\end{eqnarray} 
The asymptotic metric no longer refers to the central black hole. It
is the metric of a spacetime that contains only a uniform magnetic
field, expressed in an advanced coordinate system that is adapted to
the incoming light cones of an observer situated at $r=0$; the metric
is limited to a domain $r < a$, where $a$ is a length scale such that
$a^2 B^2 \ll 1$. The observer, of course, is fictitious, as $r=0$ is
actually occupied by the black-hole singularity; nevertheless, the
observer may be thought to exist in an unphysical extension of the
asymptotic spacetime beyond its domain of validity, $r \gg M$.       

The metric of an arbitrary spacetime in light-cone coordinates was     
thoroughly investigated in our companion paper
\cite{preston-poisson:06a}. By comparing our
Eqs.~(\ref{3.48})--(\ref{3.52}) to Eqs.~(4.9)--(4.12) of the companion
paper, we infer that the asymptotic spacetime is characterized by   
the following irreducible quantities: $\rho := B^2/(8\pi)$ is the 
mass-energy density of the magnetic field as measured by the observer 
at $r=0$, $S_{11} = S_{22} = -\frac{1}{2} S_{33} := B^2/(12\pi)$ are 
the nonvanishing components of the tracefree part of the field's
stress tensor, and $T := B^2/(8\pi)$ is the trace of the stress
tensor; these assignments are precisely what should be expected for a
uniform magnetic field. The comparison reveals also that 
${\cal E}_{11} = {\cal E}_{22} = -\frac{1}{2} {\cal E}_{33} 
:= {\cal E}$ are the nonvanishing components of the spacetime's Weyl
curvature tensor. (The irreducible quantities are all defined in our
companion paper.) The comparison therefore gives us an operational
meaning for the parameter ${\cal E}$: As was already anticipated, it
is the Weyl curvature (the tidal gravitational field) of the
asymptotic spacetime as measured by an observer comoving with the
black hole in the region $M \ll r \ll 1/B$.  

\subsection{Perturbed event horizon} 

The perturbed black-hole spacetime retains $\phi^\alpha$ as a
rotational Killing vector, and it retains $t^\alpha = (1,0,0,0)$ as a
time-translation Killing vector. This vector is timelike outside the
black hole, but it becomes null on the event horizon (which is
therefore a Killing horizon). Setting $g_{\alpha\beta} t^\alpha
t^\beta = g_{vv} = 0$ and involving Eq.~(\ref{3.43}) informs us that
the event horizon is now described by  
\begin{equation} 
r = r_{\rm horizon}(\theta) := 2M
\biggl(1 + \frac{2}{3} M^2 B^2 \sin^2\theta\biggr). 
\label{3.53} 
\end{equation} 
It is interesting to note that $r_{\rm horizon}(\theta)$ involves
$B^2$ but not ${\cal E}$. 

The horizon's intrinsic geometry is obtained by inserting
Eq.~(\ref{3.53}) into the perturbed metric. It is described by the
two-dimensional line element 
\begin{eqnarray}
\hspace*{-20pt}
ds^2_{\rm horizon} &=& 4M^2 \bigl[ 1 + M^2 (B^2 + 2{\cal E})
  \sin^2\theta \bigr]\, d\theta^2 
\nonumber \\ & & \hspace*{-30pt} \mbox{} 
+ 4M^2\sin^2\theta 
\bigl[ 1 - M^2 (B^2 + 2{\cal E}) \sin^2\theta \bigr]\, d\phi^2. 
\label{3.54} 
\end{eqnarray} 
The element of surface area on the horizon is $4M^2\sin\theta\,
d\theta d\phi$, and the integrated area is     
\begin{equation} 
A_{\rm horizon} = 16\pi M^2. 
\label{3.55} 
\end{equation} 
As was anticipated in Sec.~III A, the perturbed black hole has the
same surface area as the original Schwarzschild black hole; this
reflects its quasi-static and reversible immersion within the magnetic 
field. 

The distortion of the event horizon can be measured by the Ricci 
scalar associated with the two-dimensional metric of
Eq.~(\ref{3.54}). This is  
\begin{equation} 
R = \frac{1}{2M^2} \Bigl[ 1 + 2M^2 (B^2 + 2{\cal E}) 
(3\cos^2\theta -1) \Bigr]. 
\label{3.56}
\end{equation} 
The distortion has a quadrupolar structure. The larger concentration   
of curvature at the poles reflects the greater strength of the
magnetic field there; as was observed back in Sec.~III B, the square
of the magnetic field is given by $|\bm{B}|^2 = B^2\cos^2\theta$.  

It is interesting to note that in accordance with the zeroth law of
black-hole mechanics, the horizon's surface gravity displays no trace
of this distortion. The surface gravity $\kappa$ is defined
by the statement that on the horizon, $t^\alpha$ satisfies the
generalized form of the geodesic equation: $t^\beta \nabla_\beta
t^\alpha = \kappa t^\alpha$. A short calculation based on this
equation reveals that $\kappa = 1/(4M)$ plus terms of order $B^4$,
$B^2 {\cal E}$, and ${\cal E}^2$. The surface gravity is uniform on 
the horizon, and it keeps its unperturbed, Schwarzschild value.     

\subsection{Comparison with the Schwarzschild-Melvin solution} 

There exists an exact solution to the Einstein-Maxwell equations that
describes a nonrotating black hole immersed in Melvin's magnetic
universe \cite{melvin:64, melvin:65, thorne:65}. Known as the
Schwarzschild-Melvin solution \cite{ernst:76, ernst-wild:76,
hiscock:81}, it has a metric given by 
\begin{equation}
ds^2 = \Lambda^2 \bigl( -f\, dt^2 + f^{-1}\, d\bar{r}^2 + \bar{r}^2\,
d\theta^2 \bigr) + \Lambda^{-2} \bar{r}^2\sin^2\theta\, d\phi^2 
\label{3.57}
\end{equation}
and a vector potential given by 
\begin{equation}
A^\alpha = \frac{1}{2} B \Lambda \phi^\alpha, 
\label{3.58} 
\end{equation} 
where $\phi^\alpha := (0,0,0,1)$ is the spacetime's rotational Killing  
vector. We have $f := 1-2M/\bar{r}$ as before, and we introduce the  
function  
\begin{equation}   
\Lambda := 1 + \frac{1}{4} B^2 \bar{r}^2 \sin^2\theta. 
\label{3.59} 
\end{equation} 
This is a two-parameter family of black-hole solutions; the first
parameter is the black-hole mass $M$, and the second is the magnetic
field strength $B$.  

The solution of Eqs.~(\ref{3.57})--(\ref{3.59}) is exact, and we wish
to compare it with the perturbative solution of
Eqs.~(\ref{3.43})--(\ref{3.47}). We must first linearize the exact
solution with respect to $B^2$ and transform the coordinates from the
original system $(t,\bar{r},\theta,\phi)$ to the light-cone system
$(v,r,\theta,\phi)$. The transformation from $t$ to $v$ is the same as
for the Schwarzschild spacetime: 
\begin{equation} 
v = t + \bar{r} + 2M \ln(\bar{r}/2M - 1). 
\label{3.60} 
\end{equation} 
The transformation from $\bar{r}$ to $r$ is designed to change the
$g_{vr}$ component of the metric tensor from its current value
$\Lambda^2 \simeq 1 + \frac{1}{2} B^2 \bar{r}^2\sin^2\theta$ to the
new value of $1$. It is given by  
\begin{equation} 
r = \bar{r} \biggl[ 1 + \frac{1}{6} B^2 \bar{r}^2 \sin^2\theta
+ O(B^4) \biggr]. 
\label{3.61} 
\end{equation} 
The angular coordinates $(\theta,\phi)$ are not affected by the
transformation. 

These manipulations bring the Schwarzschild-Melvin metric to the new
form 
\begin{eqnarray} 
\hspace*{-15pt} 
g_{vv} &=& -f - \frac{1}{6} B^2 r(3r-8M)\sin^2\theta + O(B^4), 
\label{3.62} \\ 
\hspace*{-15pt} 
g_{vr} &=& 1 + O(B^4), 
\label{3.63} \\ 
\hspace*{-15pt} 
g_{v\theta} &=& -\frac{1}{3} B^2 r^3 \sin\theta\cos\theta + O(B^4), 
\label{3.64} \\ 
\hspace*{-15pt} 
g_{\theta\theta} &=& r^2 + \frac{1}{6} B^2 r^4 \sin^2\theta + O(B^4), 
\label{3.65} \\ 
\hspace*{-15pt} 
g_{\phi\phi} &=& r^2\sin^2\theta - \frac{5}{6} B^2 r^4 \sin^4\theta 
+ O(B^4).  
\label{3.66}  
\end{eqnarray} 
Comparison with Eqs.~(\ref{3.43})--(\ref{3.47}) reveals that the
solutions are identical provided that we restrict the parameter
freedom of the perturbative solution. Indeed, to get a match we
must set  
\begin{equation} 
{\cal E} = \frac{1}{2} B^2.
\label{3.67} 
\end{equation} 
The Weyl curvature of the Schwarzschild-Melvin solution is intimately
related to its magnetic field. This feature is in fact inherited from 
Melvin's pure magnetic universe, as can be inferred from reading 
Sec.~V B of our companion paper \cite{preston-poisson:06a}.    

We conclude with the following statement: While the
Schwarzschild-Melvin solution has the advantage of being an exact
solution to the Einstein-Maxwell equations, the perturbative solution
of Eqs.~(\ref{3.43})--(\ref{3.47}) has the advantage of possessing a 
larger number of parameters. The perturbative solution can therefore
represent a wider class of physical situations. In particular, it
provides the description of a magnetized black-hole spacetime in which
the tidal gravity is not directly tied to the magnetic field.         

\begin{acknowledgments} 
We thank Bernie Nickel and Bernard Whiting for many valuable
discussions. This work was supported by the Natural Sciences and
Engineering Research Council of Canada.     
\end{acknowledgments} 

\appendix 
\section*{Perturbation equations in the light-cone gauge} 

In the even-parity sector the nonvanishing perturbation fields are 
$h_{vv}(v,r)$, $j_v(v,r)$, $K(v,r)$, and $G(v,r)$. According to
Eqs.~(4.13)--(4.16) of Martel \& Poisson \cite{martel-poisson:05},
they satisfy the field equations 
\begin{widetext} 
\begin{eqnarray*} 
Q^{vv} &=& - \frac{\partial^2}{\partial r^2} K 
- \frac{2}{r} \frac{\partial}{\partial r} K, 
\\ 
Q^{vr} &=& \frac{\partial^2}{\partial v \partial r} K 
+ \frac{2}{r} \frac{\partial}{\partial v} K
- \frac{1}{r} \frac{\partial}{\partial r} h_{vv} 
+ \frac{\lambda}{2 r^2} \frac{\partial}{\partial r} j_v  
+ \frac{r-M}{r^2} \frac{\partial}{\partial r} K 
- \frac{1}{r^2} h_{vv} 
+ \frac{\lambda}{r^3} j_v
- \frac{\mu}{2r^2} K
- \frac{\mu\lambda}{4 r^2} G, 
\\ 
Q^{rr} &=& -\frac{\partial^2}{\partial v^2} K 
+ \frac{r-M}{r^2} \frac{\partial}{\partial v} K 
+ \frac{1}{r} \frac{\partial}{\partial v} h_{vv}
- \frac{\lambda}{r^2} \frac{\partial}{\partial v} j_v
- \frac{f}{r} \frac{\partial}{\partial r} h_{vv}  
+ \frac{(r-M)f}{r^2} \frac{\partial}{\partial r} K
\\ & & \mbox{}
+ \frac{\mu r + 4M}{2r^3} h_{vv}
+ \frac{\lambda(r-M)}{r^4} j_v 
- \frac{\mu f}{2r^2} K
- \frac{\mu\lambda f}{4 r^2} G, 
\\ 
Q^{v} &=&  
\frac{\partial^2}{\partial r^2} j_v 
- \frac{\partial}{\partial r} K 
- \frac{\mu}{2} \frac{\partial}{\partial r} G 
- \frac{2}{r^2} j_v, 
\\
Q^{r} &=& -\frac{\partial^2}{\partial v\partial r} j_v 
+ \frac{2}{r} \frac{\partial}{\partial v} j_v 
- \frac{\partial}{\partial v} K
- \frac{\mu}{2} \frac{\partial}{\partial v} G
- \frac{\mu f}{2} \frac{\partial}{\partial r} G
- f \frac{\partial}{\partial r} K 
+ \frac{\partial}{\partial r} h_{vv} 
- \frac{2}{r^2} j_v, 
\\
Q^\flat &=& 2 \frac{\partial^2}{\partial v \partial r} K 
+ \frac{2}{r} \frac{\partial}{\partial v} K
+ f \frac{\partial^2}{\partial r^2} K 
- \frac{\partial^2}{\partial r^2} h_{vv} 
- \frac{2}{r} \frac{\partial}{\partial r} h_{vv} 
+ \frac{\lambda}{r^2} \frac{\partial}{\partial r} j_v
+ \frac{2(r-M)}{r^2} \frac{\partial}{\partial r} K,  
\\ 
Q^\sharp &=& -2 r^2 \frac{\partial^2}{\partial v\partial r} G 
- 2 r \frac{\partial}{\partial v} G
- r^2 f \frac{\partial^2}{\partial r^2} G
- 2 (r-M) \frac{\partial}{\partial r} G
+ 2 \frac{\partial}{\partial r} j_v,  
\end{eqnarray*} 
\end{widetext}
where $\lambda := l(l+1) = \mu + 2$ and $\mu := (l-1)(l+2) 
= \lambda - 2$. According to Eqs.~(4.17)--(4.20) of Martel \& Poisson
\cite{martel-poisson:05}, the source terms are 
\begin{eqnarray*}  
Q^{ab} &=& 8\pi \int T^{ab} \bar{Y}^{lm}\, d\Omega, 
\\
Q^a &=& \frac{16\pi r^2}{l(l+1)} \int T^{aB} \bar{Y}^{lm}_B\, d\Omega, 
\\ 
Q^\flat &=& 8\pi r^2 \int T^{AB} \Omega_{AB} \bar{Y}^{lm}\, d\Omega, 
\\ 
Q^\sharp &=& \frac{32\pi r^4}{(l-1)l(l+1)(l+2)} \int T^{AB}
\bar{Y}^{lm}_{AB}\, d\Omega,  
\end{eqnarray*} 
where $x^a = (v,r)$. The perturbation equations are not all
independent; they are linked by the Bianchi identities
\begin{eqnarray*}  
0 &=& \frac{\partial}{\partial v} Q^{vv} 
+ \frac{\partial}{\partial r} Q^{vr} 
+ \frac{M}{r^2} Q^{vv} 
\\ & & \mbox{} 
+ \frac{2}{r} Q^{vr} 
- \frac{\lambda}{2r^2} Q^v - \frac{1}{r} Q^\flat, \\ 
0 &=& \frac{\partial}{\partial v} Q^{vr} 
+ \frac{\partial}{\partial r} Q^{rr} 
+ \frac{Mf}{r^2} Q^{vv} 
- \frac{2M}{r^2} Q^{vr} 
\\ & & \mbox{} 
+ \frac{2}{r} Q^{rr}
- \frac{\lambda}{2r^2} Q^r 
- \frac{f}{r} Q^\flat, \\ 
0 &=& \frac{\partial}{\partial v} Q^v 
+ \frac{\partial}{\partial r} Q^r 
+ \frac{2}{r} Q^r + Q^\flat - \frac{\mu}{2r^2} Q^\sharp. 
\end{eqnarray*}  
When $l=0$ the only nonvanishing perturbation fields are $h_{vv}$ and
$K$, and the only relevant equations are those involving $Q^{ab}$ and
$Q^\flat$. When $l=1$ the only nonvanishing perturbation fields are
$h_{vv}$, $j_v$, and $K$, and the only relevant equations are those
involving $Q^{ab}$, $Q^a$, and $Q^\flat$. 

In the odd-parity sector the nonvanishing perturbation fields are 
$h_{v}(v,r)$ and $h_2(v,r)$. According to Eq.~(5.8) and (5.9) of
Martel \& Poisson \cite{martel-poisson:05}, they satisfy the field
equations  
\begin{widetext} 
\begin{eqnarray*} 
P^v &=& \frac{\partial^2}{\partial r^2} h_v
- \frac{\mu}{2 r^2} \frac{\partial}{\partial r} h_2 
- \frac{2}{r^2} h_v 
+ \frac{\mu}{r^3} h_2, 
\\ 
P^r &=& -\frac{\partial^2}{\partial v \partial r} h_v 
+ \frac{2}{r} \frac{\partial}{\partial v} h_v 
- \frac{\mu}{2 r^2} \frac{\partial}{\partial v} h_2 
- \frac{\mu f}{2 r^2} \frac{\partial}{\partial r} h_2 
+ \frac{\mu}{r^2} h_v
+ \frac{\mu f}{r^3} h_2,
\\
P &=& -\frac{\partial^2}{\partial v\partial r} h_2 
+ \frac{1}{r} \frac{\partial}{\partial v} h_2 
- \frac{f}{2} \frac{\partial^2}{\partial r^2} h_2  
+ \frac{r-3M}{r^2} \frac{\partial}{\partial r} h_2  
+ \frac{\partial}{\partial r} h_v
- \frac{r-4M}{r^3} h_2,
\end{eqnarray*} 
\end{widetext} 
where $\lambda := l(l+1) = \mu + 2$ and $\mu := (l-1)(l+2) 
= \lambda - 2$. According to Eqs.~(5.10) and (5.11) of Martel \&
Poisson \cite{martel-poisson:05}, the source terms are  
\begin{eqnarray*}  
P^a &=& \frac{16\pi r^2}{l(l+1)} \int T^{aB} \bar{X}^{lm}_B\, d\Omega,  
\\ 
P &=& \frac{16\pi r^4}{(l-1)l(l+1)(l+2)} \int T^{AB}
\bar{X}^{lm}_{AB}\, d\Omega. 
\end{eqnarray*} 
The perturbation equations are not all independent; they are linked by
the Bianchi identity 
\[
0 = \frac{\partial}{\partial v} P^v + \frac{\partial}{\partial r} P^r  
+ \frac{2}{r} P^r - \frac{\mu}{r^2} P. 
\]
When $l=1$ the only nonvanishing perturbation field is $h_v$, and the
only relevant equations are those involving $P^a$.

\bibliography{../bib/master}

\begin{thebibliography}{10}
\expandafter\ifx\csname url\endcsname\relax
  \def\url#1{{\tt #1}}\fi
\expandafter\ifx\csname urlprefix\endcsname\relax\def\urlprefix{URL }\fi

\bibitem{chandrasekhar:83}
S.~Chandrasekhar, {\em The mathematical theory of black holes\/} (Clarendon
  Press, Oxford, England, 1983).

\bibitem{regge-wheeler:57}
T.~Regge and J.~A. Wheeler, {\em Stability of a Schwarzschild singularity\/},
  Phys. Rev. {\bf 108}, 1063 (1957).

\bibitem{vishveshwara:70}
C.~V. Vishveshwara, {\em Stability of the Schwarzschild metric\/}, Phys. Rev. D
  {\bf 1}, 2870 (1970).

\bibitem{zerilli:70}
F.~J. Zerilli, {\em Gravitational field of a particle falling in a
  Schwarzschild geometry analyzed in tensor harmonics\/}, Phys. Rev. D {\bf 2},
  2141 (1970).

\bibitem{poisson:04b}
E.~Poisson, {\em The motion of point particles in curved spacetime\/}, Living
  Rev. Relativity {\bf 7} (2004), 6. [Online article]: cited on \today,
  http://www.livingreviews.org/lrr-2004-6.

\bibitem{detweiler:05}
S.~Detweiler, {\em Perspective on gravitational self-force analyses\/}, Class.
  Quantum Grav. {\bf 22}, S681 (2005), arXiv:gr-qc/0501004.

\bibitem{hikida-etal:05}
W.~Hikida, H.~Nakano, and M.~Sasaki, {\em Self-force Regularization in the
  Schwarzschild Spacetime\/}, Class. Quantum Grav {\bf 22}, S753 (2005),
  arXiv:gr-qc/0411150.

\bibitem{drasco-etal:05}
S.~Drasco, E.~E. Flanagan, and S.~A. Hughes, {\em Computing inspirals in Kerr
  in the adiabatic regime. I. The scalar case\/}, Class. Quantum Grav. {\bf
  22}, S801 (2005), arXiv:gr-qc/0505075.

\bibitem{gleiser-etal:00}
R.~Gleiser, C.~Nicasio, R.~Price, and J.~Pullin, {\em Gravitational radiation
  from Schwarzschild black holes: The second order perturbation formalism\/},
  Phys. Rept. {\bf 325}, 41 (2000).

\bibitem{alvi:00}
K.~Alvi, {\em Approximate binary-black-hole metric\/}, Phys. Rev. D {\bf 61},
  124013 (2000), arXiv:gr-qc/9912113.

\bibitem{alvi:01}
K.~Alvi, {\em Energy and angular momentum flow into a black hole in a
  binary\/}, Phys. Rev. D {\bf 64}, 104020 (2001), arXiv:gr-qc/0107080.

\bibitem{alvi:03}
K.~Alvi, {\em Note on ingoing coordinates for binary black holes\/}, Phys. Rev.
  D {\bf 67}, 104006 (2003), arXiv:gr-qc/0302061.

\bibitem{yunes-etal:06}
N.~Yunes, W.~Tichy, B.~J. Owen, and B.~Bruegmann, {\em Binary black hole
  initial data from matched asymptotic expansions\/} (2005),
  arXiv:gr-qc/0503011.

\bibitem{yunes-gonzalez:06}
N.~Yunes and J.~Gonzalez, {\em Metric of a tidally perturbed spinning black
  hole\/}, Phys. Rev. D {\bf 73}, 024010 (2006).

\bibitem{moncrief:74}
V.~Moncrief, {\em Gravitational perturbations of spherically symmetric systems.
  I. The exterior problem\/}, Ann. Phys. (N.Y.) {\bf 88}, 323 (1974).

\bibitem{gerlach-sengupta:79}
U.~H. Gerlach and U.~K. Sengupta, {\em Gauge-invariant perturbations on most
  general spherically symmetric space-times\/}, Phys. Rev. D {\bf 19}, 2268
  (1979).

\bibitem{sarbach-tiglio:01}
O.~Sarbach and M.~Tiglio, {\em Gauge-invariant perturbations of Schwarzschild
  black holes in horizon-penetrating coordinates\/}, Phys. Rev. D {\bf 64},
  084016 (2001), arXiv:gr-qc/0104061.

\bibitem{gundlach-martingarcia:00}
C.~Gundlach and J.~M. Martin-Garcia, {\em Gauge-invariant and
  coordinate-independent perturbations of stellar collapse: The interior\/},
  Phys. Rev. D {\bf 61}, 084024 (2000), arXiv:gr-qc/9906068.

\bibitem{clarkson-barrett:03}
C.~Clarkson and R.~Barrett, {\em Covariant perturbations of Schwarzschild black
  holes\/}, Class. Quantum Grav. {\bf 20}, 3855 (2003), arXiv:gr-qc/0209051.

\bibitem{nagar-rezzolla:05}
A.~Nagar and L.~Rezzolla, {\em Gauge-invariant non-spherical metric
  perturbations of Schwarzschild spacetime\/}, Class. Quantum Grav. {\bf 22},
  R167 (2005), arXiv:gr-qc/0502064.

\bibitem{martel-poisson:05}
K.~Martel and E.~Poisson, {\em Gravitational perturbations of the Schwarzschild
  spacetime: A practical covariant and gauge-invariant formalism\/}, Phys. Rev.
  D {\bf 71}, 104003 (2005), arXiv:gr-qc/0502028.

\bibitem{ernst:76}
F.~J. Ernst, {\em Black holes in a magnetic universe\/}, J. Math. Phys. {\bf
  17}, 54 (1976).

\bibitem{ernst-wild:76}
F.~J. Ernst and W.~J. Wild, {\em Kerr black holes in a magnetic universe\/}, J.
  Math. Phys. {\bf 17}, 182 (1976).

\bibitem{hiscock:81}
W.~A. Hiscock, {\em On black holes in magnetic universes\/}, J. Math. Phys {\bf
  22}, 1828 (1981).

\bibitem{preston-poisson:06a}
B.~Preston and E.~Poisson, {\em Light-cone coordinates based at a geodesic
  world line\/} (2006), to appear.

\bibitem{MTW:73}
C.~W. Misner, K.~S. Thorne, and J.~A. Wheeler, {\em Gravitation\/} (Freeman,
  San Francisco, 1973).

\bibitem{chrzanowski:75a}
P.~L. Chrzanowski, {\em Vector potential and metric perturbations of a rotating
  black hole\/}, Phys. Rev. D {\bf 11}, 2042 (1975).

\bibitem{wald:78}
R.~M. Wald, {\em Construction of solutions of gravitational, electromagnetic,
  or other perturbation equations from solutions of decoupled equations\/},
  Phys. Rev. Lett. {\bf 41}, 203 (1978).

\bibitem{lousto-whiting:02}
C.~O. Lousto and B.~F. Whiting, {\em Reconstruction of black hole metric
  perturbations from the Weyl curvature\/}, Phys. Rev. D {\bf 66}, 024026
  (2002), arXiv:gr-qc/0203061.

\bibitem{ori:03}
A.~Ori, {\em Reconstruction of inhomogeneous metric perturbations and
  electromagnetic four-potential in Kerr spacetime\/}, Phys. Rev. D {\bf 67},
  124010 (2003), arXiv:gr-qc/0207045.

\bibitem{price:05}
L.~Price, {\em Towards metric perturbations of Kerr\/} (2005),
  http://www.sstd.rl.ac.uk/capra/Price.pdf. Oral presentation at the 8th Capra
  Meeting on Radiation Reaction, July 11--14, 2005.

\bibitem{price-etal:06}
L.~Price, K.~Shankar, and B.~F. Whiting, {\em On the existence of radiation
  gauges in Petrov type II space-times\/} (2006), to appear.

\bibitem{whiting-price:05}
B.~F. Whiting and L.~R. Price, {\em Metric reconstruction from Weyl scalars\/},
  Class. Quantum Grav. {\bf 22}, S589 (2005).

\bibitem{wald:74}
R.~M. Wald, {\em Black hole in a uniform magnetic field\/}, Phys. Rev. D {\bf
  10}, 1680 (1974).

\bibitem{melvin:64}
M.~A. Melvin, {\em Pure magnetic and electric geons\/}, Phys. Letters {\bf 8},
  65 (1964).

\bibitem{melvin:65}
M.~A. Melvin, {\em Dynamics of cylindrical electromagnetic universes\/}, Phys.
  Rev. {\bf 139}, B225 (1965).

\bibitem{thorne:65}
K.~S. Thorne, {\em Absolute stability of Melvin's magnetic universe\/}, Phys.
  Rev. {\bf 139}, B244 (1965).

\end{thebibliography}
\end{document}